**An Algebraic Method for the Analytical Solutions of the Klein-Gordon equation for any angular momentum for some diatomic potentials**

**Huseyin Akcay** [1,a] **and Ramazan Sever** [2,b]

[1] Faculty of Engineering, Başkent University, Baglıca Campus, Ankara, Turkey

[2] Department of Physics, Faculty of Arts and Sciences, Middle East Technical University, 06531 Ankara, Turkey

**Abstract**

Analytical solutions of the Klein-Gordon equation are obtained by reducing the radial part of the wave equation to a standard form of a second order differential equation. Differential equations of this standard form are solvable in terms of hypergeometric functions and we give an algebraic formulation for the bound state wave functions and for the energy eigenvalues. This formulation is applied for the solutions of the Klein-Gordon equation with some diatomic potentials.

[a] E-mail: akcay@baskent.edu.tr

[b] E-mail: sever@metu.edu.tr

### 1  Introduction

Dirac equation and Klein-Gordon ( K-G) equations are used for the description of particle dynamics in relativistic quantum mechanics. Therefore, the search for the solutions of these equations has been an important research area and several methods were used to obtain solutions of these equations with various physical potentials. These methods include the Asymptotic Iteration method [1-4], the Nikiforov-Uvarov method [5-7], Supersymmetric method [8-12], Variational method [13,14], the Large-N Expansion method [15], Exact Quantization Rule [16-18]. The vector and scalar potential couplings are introduced into the relativistic wave equations by using the four-vector linear momentum and the mass terms which are present in the free form of these equations. Recently, there has been an interest in finding solutions of these equations with mixtures of scalar and vector potentials [4,19]. This is important for the studies of pseudospin symmetry in nuclear physics [20].
Unfortunately, these wave equations with some important potentials can be solved only for states with zero angular momentum [21-23]. In this work we study any-$\ell$ state solutions for these potentials using appropriate approximations [24-27] for the centrifugal term .

In this paper we use an algebraic method to find the analytical solutions of Klein-Gordon equation. We have used this method for the Schrödinger and Dirac equations and obtained analytical solutions of these equations for a large set of potentials [28,29]. Here we consider a set of physical potentials including Hulthen Potential [21,30-33], Coulomb potential [33,34], Mie Potential [35-40], Woods-Saxon Potential [41-46], Kratzer-Fues Potential [47,48], Non-central Potential [49-52], Pöschl-Teller Potential 53-59]. We study the solutions of the wave equations with these potentials and we include both scalar and vector interactions in our analysis. As stated before, the K-G equation with some of these potentials have been investigated using different methods . We compare our results with the results of the existing literature. In the applications we have given some details of this comparison for the Coulomb Potential, the Kratzer-Fues Potential, the generalized Hulthén potential and for the Generalized Pöschl- Teller potential.

The organization of the paper is as follows. In the next section we give a brief outline of our scheme. In section 3 we give several application of the scheme for the solution of the Klein-Gordon equation. We obtain the wave functions and write down equation for the energy eigenvalues for each application. Finally we summarize our conclusions in section 4.

## 2  A Brief Review of the Method

For many problems of quantum mechanics that can be solved analytically the radial wave equation reduces to the following second order differential equation

$$\frac{d^2\psi}{ds^2} + \frac{(c_1+c_2 s)}{s(1+c_3 s)}\frac{d\psi}{ds} + \frac{1}{s^2(1+c_3 s)^2}[-\Lambda_1 s^2 + \Lambda_2 s - \Lambda_3]\psi = 0 \tag{1}$$

where $c_i$ and $\Lambda_i$ are some constants. We have shown in Ref. 28 that, differential equations of this parametric form have bound state solutions. When the parameter $c_3$ is not zero the solutions are given in terms of Jacobi polynomials (the hypergeometric functions ), when $c_3$ is zero they are in given in terms of Laguerre polynomials (the confluent hypergeometric functions ). In this section we give a short summary of the scheme. The details of the derivations are given in reference 28 . When $c_3$ is not zero the solutions can are written as

$$\Psi(s) = (1+c_3 s)^{-p_1} s^{q_1} y(s) \tag{2}$$

where $q_1$ and $p_1$ are some arbitrary constants and y is a new function to be determined. When we insert this function into Eq. (1) we find

$$s(1+c_3 s)\frac{d^2 y(s)}{ds^2} + A(s)\frac{dy}{ds} + B(s) y(s) = 0 \tag{3a}$$

where

$$A(s) = 2q_1(1+c_3 s) - 2p_1 c_3 s + c_1 + c_2 s \tag{3b}$$

$$B(s) = q_1(q_1-1)\frac{(1+c_3 s)}{s} - 2p_1 q_1 c_3 + \frac{p_1(p_1+1)c_3^2}{(1+c_3 s)}s \\ + \frac{(c_1+c_2 s)}{s(1+c_3 s)}[q_1 + (q_1-p_1)c_3 s] + \frac{(-\Lambda_1 s^2 + \Lambda_2 s - \Lambda_3)}{s(1+c_3 s)} \tag{3c}$$

When we define a new variable $z = 1 + 2c_3 s$ and write Eq. (3a) in terms of this new variable we obtain

$$(1-z^2)^2\frac{d^2 y}{dz^2} + (1-z^2)[\beta - \alpha - (\alpha+\beta+2)z]\frac{dy}{dz} + R(z) y = 0 \tag{3d}$$

where

$$\alpha = 2q_1 + c_1 - 1, \quad \beta = -2p_1 - c_1 + \frac{c_2}{c_3} - 1, \quad R(z) = r_1 z^2 + r_2 z + r_3 \tag{4}$$

The coefficients of the polynomial R contain only the parameters $c_i, \Lambda_i, q_1$ and $p_1$. In Ref. 28 the details of these calculations are given and it is shown that in order to have bound state solutions the following set of conditions must be satisfied

$$q_1 = (\frac{1-c_1}{2}) \pm \sqrt{(\frac{1-c_1}{2})^2 + \Lambda_3} \tag{5}$$

$$p_1 = \frac{D}{2} \pm \sqrt{(\frac{D}{2})^2 + H} \tag{6}$$

where

$$D = \frac{c_2}{c_3} - c_1 - 1, \quad H = \frac{\Lambda_1}{c_3^2} + \frac{\Lambda_2}{c_3} + \Lambda_3 \tag{7}$$

Using these equations for $q_1$ and $p_1$ one can rewrite Eq.(3b) as

$$(1-z^2)\frac{d^2 y}{dz^2} + [\beta - \alpha - (\alpha + \beta + 2)z]\frac{dy}{dz} + n(n+\alpha+\beta+1)y = 0 \tag{8}$$

where n is a positive integer. The last equation is the well-known Jacobi's differential equation and its solutions are the Jacobi polynomials $P_n^{(\alpha,\beta)}(z)$. Thus for the case $c_3 \neq 0$ the wave functions are

$$\Psi(s) = (1+c_3 s)^{-p_1} s^{q_1} P_n^{(\alpha,\beta)}(1+2c_3 s). \tag{9}$$

In addition one gets the following equation which gives the energy eigenvalues

$$(q_1 - p_1)^2 + (\frac{c_2}{c_3} + 2n - 1)(q_1 - p_1) + n(n + \frac{c_2}{c_3} - 1) = \frac{\Lambda_1}{c_3^2}. \tag{10}$$

A similar analysis can be done for the $c_3 = 0$ case. In this case Eq. (1) takes the following form

$$\frac{d^2 \psi}{ds^2} = \frac{(c_1 + c_2 s)}{s}\frac{d\psi}{ds} + \frac{1}{s^2}(-\Lambda_1 s^2 + \Lambda_2 s - \Lambda_3)\psi = 0 \tag{11}$$

and its asymptotic form suggests the following factorization

$$\psi(s) = \exp(-p_2 s) s^{q_2} y_1(s). \tag{12}$$

When we insert this into Eq. (11) we get

$$\frac{d^2 y_1}{ds^2} + [2\frac{q_2}{s} - 2p_2 + \frac{(c_1 + c_2 s)}{s}]\frac{dy_1}{ds} + [\frac{q_2(q_2 - 1)}{s^2} - 2\frac{q_2 p_2}{s} + p_1^2$$
$$+ \frac{(c_1 + c_2 s)}{s}(\frac{q_2}{s} - p_2) + (-\Lambda_1 s^2 + \Lambda_2 s - \Lambda_3)]y_1 = 0 \tag{13}$$

Writing this in terms of a new variable $z = (2p_2 - c_2)s$ and rearranging the terms it becomes

$$z^2 \frac{d^2 y_1}{dz^2} + (k+1-z)z\frac{dy_1}{dz} + A_1(z)y_1 = 0$$
$$k = c_1 + 2q_2 - 1 \tag{14}$$

Here $A_1(z)$ is a second order polynomial in the variable z and its coefficients are given in reference 28. We do not write these coefficients here since they will take large space. The condition on the physical bound states requires the following relations

$$q_2 = \frac{(1-c_1)}{2} + \sqrt{(\frac{1-c_1}{2})^2 + \Lambda_3} \qquad (15)$$

$$p_2 = \frac{c_2}{2} + \sqrt{(\frac{c_2}{2})^2 + \Lambda_1} \; . \qquad (16)$$

And equation (14) takes the following form

$$z\frac{d^2 y_1}{dz^2} + (k+1-z)\frac{dy_1}{dz} + ny_1 = 0 \qquad (17)$$

where n is a positive integer. From the conditions on the parameters it also follows that

$$c_1 p_2 - q_2(c_2 - 2p_2) - \Lambda_2 = n(c_2 - 2p_2) \qquad (18)$$

which is an equation for the energy eigenvalues. Eq. (17) is the Laguerre's associated differential equation and its solutions are the associated Laguerrepolynomials $L_n^k(z)$. Thus the wave function can be written down as

$$\psi(s) = \exp(-p_2 s) s^{q_2} L_n^k([2p_2 - c_2]s) \; . \qquad (19)$$

In the following sections we give several applications of these results.

**1   Applications**

The time independent Klein-Gordon equation with a vector potential V and a scalar potential S can be written as  ($\hbar = 1, c = 1$)

$$[-\nabla^2 + (m+S)^2]\psi(\vec{r}) = (E-V)^2 \psi(\vec{r}) \; . \qquad (20)$$

When the potentials are spherically symmetric, the wave functions are written as $\psi(\vec{r}) = R(r)Y_{\ell m}(\theta,\phi)$ Eq.(20) gives

$$\frac{d^2 R}{dr^2} + \frac{2}{r}\frac{dR}{dr} + \{(E-V)^2 - \frac{\ell(\ell+1)}{r^2} - (m+S)^2\}R = 0 \qquad (21)$$

for the radial part of the wave functions.

We start with the potentials which lead to a radial equation given by Eq. (11). That is applications for $c_3 = 0$ with $S \neq V$.

Case 1:  Coulomb Potential

Let us take the vector potential as $V(r) = -Z\alpha/r$ and assume that the scalar potential is zero. This can be considered as the potential for the pionic atom. This allows us to write Eq. (21) as

$$\frac{d^2 R}{dr^2} + \frac{2}{r}\frac{dR}{dr} + \frac{1}{r^2}[-\ell(\ell+1) + (Z\alpha)^2 + (E^2 - m^2)r^2 + 2EZ\alpha r]R = 0 \qquad (22)$$

Comparing this equation with Eq. (1) we get $c_1 = 2, c_2 = 0, c_3 = 0$, $\Lambda_1 = (m^2 - E^2)$, $\Lambda_2 = 2EZ\alpha$, $\Lambda_3 = \ell(\ell+1) - (Z\alpha)^2$ Thus we must use Eqs. (15) and (16) for the calculation of $q_2$ and $p_2$, respectively and also Eq. (14) for calculating k. We find

$$q_2 = -\frac{1}{2} + \sqrt{(\ell+\frac{1}{2})^2 - (Z\alpha)^2}, \quad p_2 = \sqrt{\Lambda_1}, k = 2q_2 + 1 \tag{23}$$

Now we can determine the energy eigenvalues using Eq. (18) which gives $p_2(2 + 2q_2 + 2n) = \Lambda_2$. Replacing the parameters and solving this equation for E we obtain

$$E_{n\ell} = \pm m\{1 + \frac{(Z\alpha)^2}{[n + \frac{1}{2}\sqrt{(\ell+\frac{1}{2})^2 - (Z\alpha)^2}\,]^2}\}^{-\frac{1}{2}} \tag{24}$$

We can choose the negative sign for the bound states. These are the energy levels for a pionic atom. In Ref. (33) K-G equation with Coulomb Potential is solved with a different method. The energy spectrum given by Eq. (24) agrees exactly with their result. It is also solved with the same potential in D dimensions in Ref. (54). Our results agree with this reference for D=3. The wave functions corresponding to Eq. (24) are given by Eq.(19) and can be written as

$$R_{n\ell}(r) = N \exp(-\sqrt{(m^2 - E^2)}r) r^{(\sqrt{(\ell+\frac{1}{2})^2 - (Z\alpha)^2} - \frac{1}{2})} L_n^{2(\sqrt{(\ell+\frac{1}{2})^2 - (Z\alpha)^2})}(2\sqrt{(m^2 - E^2)}r), |E| \leq m \tag{25}$$

where N is a constant. This wave function agrees with the wave function of reference (33). If we take a Coulomb-like scalar potential S we can find the analytical solutions using the same method.

Applications with $c_3 = 0$ and $S = V$:

Case 2: Mie Potential

The Mie potential is given by [35-40]

$$V(r) = V_0[\frac{1}{2}(\frac{a}{r})^2 - \frac{a}{r}]. \tag{26}$$

Assuming that the vector and scalar potentials are equal in Eq. (21) which can be written as

$$\frac{d^2R}{dr^2} + \frac{2}{r}\frac{dR}{dr} + \frac{1}{r^2}\{(E^2 - m^2)r^2 + 2(E+m)aV_0 r - \ell(\ell+1) - (E+m)a^2 V_0\}R = 0 \tag{27}$$

Comparing this with Eq. (11) we can write the parameters of this equation as follows

$$c_1 = 2, c_2 = 0, c_3 = 0, \quad \Lambda_1 = (m^2 - E^2), \Lambda_2 = 2(E+m)aV_0, \Lambda_3 = \ell(\ell+1) + (E+m)a^2 V_0. \tag{28}$$

Inserting these parameters into Eqs. (15) and (16) we find $q_2 = -\frac{1}{2} + \frac{1}{2}\sqrt{1+4\Lambda_3}$, $p_2 = \sqrt{\Lambda_1}$, and using Eq. (14) we find $k = \sqrt{1+4\Lambda_3}$. Using these parameters in Eq.(18) we find following equation for the energy eigenvalues

$$E_{n\ell}^2 = m^2(1 - \frac{4(\frac{E_{n\ell}}{m}+1)a^2V_0^2}{[2n+1+\sqrt{(2\ell+1)^2 + 4(E_{n\ell}+m)a^2V_0}]^2}) \quad (29)$$

The corresponding wave functions are obtained from Eq. (19) as

$$R(r) = N\exp(-\sqrt{\Lambda_1}r)r^{\frac{1}{2}(\sqrt{1+4\Lambda_3})-1)}L_n^{\sqrt{1+4\Lambda_3}}(2\sqrt{\Lambda_1}r) \quad (30)$$

Case 3: Kratzer-Fues Potential

The Kratzer-Fues potential is [47,48] $V(r) = V_e(r-r_e)^2/r^2$ and we will assume that S=V. Then Eq.(21) takes the following form

$$\frac{d^2R}{dr^2} + \frac{2}{r}\frac{dR}{dr} + \frac{1}{r^2}\{-\ell(\ell+1) - [2(E+m)V_e - (E^2-m^2)]r^2 + [4(E+m)V_e r_e]r - 2(E+m)V_e r_e^2\}R = 0 \quad (31)$$

Using Eqs.(11,15,16) we find $c_1 = 2$, $c_2 = 0$, $c_3 = 0$, $\Lambda_1 = 2(E+m)V_e - (E^2-m^2)$, $\Lambda_2 = 4(E+m)V_e r_e$, $\Lambda_3 = \ell(\ell+1) + (2(E+m)V_e r_e^2$. $q_2 = \frac{1}{2}(\sqrt{1+\Lambda_3}-1)$, $p_2 = \sqrt{\Lambda_1}$. Thus the equation for the energy eigenvalues given by Eq.(18) reduces to $(2n+2+q_2)p_2 = \Lambda_2$ which can be written as

$$E_{n\ell}^2 = m^2 - \frac{[4(E_{n\ell}+m)V_e r_e]^2}{[2n+1+\sqrt{1+4\ell(\ell+1)+8(E_{n\ell}+m)V_e r_e^2}]^2} + 2(E_{n\ell}+m)V_e \quad . \quad (32)$$

The wave functions are written using Eq.(19) as

$$R_n(r) = N\exp(-\sqrt{\Lambda_1}r)r^{\frac{1}{2}(\sqrt{1+4\Lambda_3}-1)}L_n^{(\sqrt{1+4\Lambda_3})}(2\sqrt{\Lambda_1}r) \quad . \quad (33)$$

In Ref. (48) K-G equation is solved with the method of Riccati equation. They obtain the energy levels for $V = S = -a/r + b/r^2$. Their energy levels agree with the levels given by Eq. (32). Note that we have an extra constant term in our potential which shows itself in Eq. (32)

Case 4: The non-central potential

We write the non-central potential [49-52] as

$$V = \frac{\alpha}{r} + \frac{\beta}{r^2 \cos^2(\theta)} \quad . \tag{34}$$

Assuming that the scalar potential is equal to vector potential, K-G equation given by Eq.(20) is separable and writing the wave function as $\psi = R(r)\varphi(\theta,\phi)$ we get

$$\frac{d^2 R}{dr^2} + \frac{2}{r}\frac{dR}{dr} + \frac{1}{r^2}[(E^2 - m^2)r^2 - 2(E+m)\alpha r - \lambda]R = 0 \tag{35}$$

for the radial wave function. Here $\lambda$ is a constant which can be written in terms of the angular quantum number $\ell$. From Eq. (11) we find $c_1 = 2$, $c_2 = 0$, $c_3 = 0$. $\Lambda_1 = (m^2 - E^2)$, $\Lambda_2 = -2(E+m)\alpha$, $\Lambda_3 = \lambda$. $\Lambda_3 = \ell(\ell+1) - (Z\alpha)^2$. Thus we must use Eqs.(15) and (16) for the calculation of $q_2$ and $p_2$, respectively and also Eq.(14) for calculating k. We find

$p_2 = \sqrt{\Lambda_1}$, $q_2 = \frac{1}{2}(\sqrt{1+4\lambda} - 1)$, $k = \sqrt{1+4\lambda}$. The energy eigenvalues satisfy Eq.(18) which gives

$$E_n^2 = m^2 - \frac{4(E_n+m)^2 \alpha^2}{(2n+1+\sqrt{1+4\lambda})^2} \quad . \tag{36}$$

The corresponding wave functions are

$$R_n(r) = N \exp(-\sqrt{\Lambda_1} r) r^{\frac{1}{2}(\sqrt{1+4\lambda}-1)} L_n^{\sqrt{1+4\lambda}}(2\sqrt{\Lambda_1} r) \quad . \tag{37}$$

Applications with $c_3 \neq 0$ and $S \neq V$:

Case 5: The generalized Hulthén potential:

This potential is defined as [21,30-33]

$$V(r) = -V_0 \frac{\exp(-\delta r)}{[1 - q\exp(-\delta r)]} \tag{38}$$

where $V_0$ is the depth of the potential, $\delta$ is the screening parameter and $q \neq 0$ is the deformation parameter. For this potential we first write equation (21) as

$$\frac{d^2 u_n(r)}{dr^2} + [-(m+S)^2 + (E-V)^2 - \frac{\ell(\ell+1)}{r^2}]u_n(r) = 0 \tag{39}$$

where $u_n(r) = R_n(r)r$. We first consider the zero angular momentum ($\ell = 0$) case and assume a scalar potential of the form $S(r) = -S_0 \exp(-\delta r)/[1 - q\exp(-\delta r)]$. It is more convenient to use a new variable s which is defined as $s = 1/[1 - q\exp(-\delta r)]$, $(r \in [0,\infty), s \in [(1-q)^{-1}, 1))$. Using this variable with Eq.(39) we obtain

$$\frac{d^2 u}{ds^2} + \frac{(1-2s)}{s(1-s)}\frac{du}{ds} + \frac{1}{s^2(1-s)^2}[-\Lambda_1 s^2 + \Lambda_2 s - \Lambda_3]u = 0 \tag{40}$$

where

$$\Lambda_1 = \frac{(S_0^2 - V_0^2)}{\delta^2 q^2}, \quad \Lambda_2 = 2\frac{(S_0^2 - V_0^2)}{\delta^2 q^2} + \frac{2mS_0 + 2EV_0}{\delta^2 q}$$

$$\Lambda_3 = \frac{(S_0^2 - V_0^2)}{\delta^2 q^2} + \frac{2mS_0 + 2EV_0}{\delta^2 q} - \frac{E^2 - m^2}{\delta^2}. \tag{41}$$

Eq. (40) has the form given in Eq.(1). Therefore we can write the parameters as $c_1 = 1$, $c_2 = -2$, $c_3 = -1$. We use Eqs .(5) and (6) for the calculation of $q_1$ and $p_1$ respectively and obtain $q_1 = \pm\sqrt{\Lambda_3}$, $p_1 = \pm\sqrt{(m^2 - E^2)/\delta^2}$. Inserting these into Eq.(10) we find the following equation for the energy eigenvalues

$$\{(2n+1) + 2\sqrt{\frac{(S_0^2 - V_0^2)}{\delta^2 q^2} + \frac{2mS_0 + 2E_n V_0}{\delta^2 q} - \frac{E_n^2 - m^2}{\delta^2}} - 2\frac{\sqrt{(m^2 - E_n^2)}}{\delta}\}^2 = 1 + 4\frac{(S_0^2 - V_0^2)}{\delta^2 q^2}. \tag{42}$$

The corresponding wave functions are given by Eq. (9) as

$$u_{n\ell}(s) = C(1-s)^{-\sqrt{H}} s^{\sqrt{\Lambda_3}} P_n^{(2\sqrt{\Lambda_3}, -2\sqrt{H})}(1-2s) \tag{43}$$

where H is given by $H = (m^2 - E^2)/\delta^2$. When we consider an arbitrary angular momentum $\ell$ we have to add the angular momentum term $\ell(\ell+1)/r^2$ to Eq. (40). In order to find analytical solutions we will use the approximation [24-27] $1/r^2 \approx \delta^2 \exp(-\delta r)/[1-\exp(-\delta r)]^2$ for the centrifugal term. When we choose $q = 1$ and include the contribution of angular momentum term in our calculation we find that only the parameters $\Lambda_i$, in Eq.(40), changes to some new parameters $\Gamma_i$. These new parameters are given by $\Gamma_1 = \Lambda_1 + \ell(\ell+1)$, $\Gamma_2 = \Lambda_2 + \ell(\ell+1)$, $\Gamma_3 = \Lambda_3$. Therefore, we have to replace the parameters $\Lambda_i$ with $\Gamma_i$ in the equations for the energy levels and wave functions. When we compare our result with Ref. (31) where the K-G equation is solved in D-dimensions. We find that the wave functions are the same. For the energy eigenvalues they use a different set of parameters.

Case 6:  Deformed Woods-Saxon Potential

This potential is given by [41-46]

$$V = -V_0 \frac{1}{[1 + q\exp(vx)]} \tag{44}$$

where $x = r - R$, $V_0$, $q$ and $v = 1/a$ are potential parameters. Assuming that the scalar potential has the same form as $V$ with $V_0$ replaced by $S_0$ we will solve the zero angular momentum K-G equation. We define a new variable $s = 1/[1 + q\exp(vx)]$ and write the K-G equation in terms of this variable. We obtain

$$\frac{d^2 u}{ds^2} + \frac{(1-2s)}{s(1-s)}\frac{du}{ds} + \frac{1}{s^2(1-s)^2}[-\Lambda_1 s^2 + \Lambda_2 s + \Lambda_3]u = 0 \tag{45}$$

where $\quad \Lambda_1 = (S_0^2 - V_0^2)a^2, \quad \Lambda_2 = 2a^2(EV_0 + mS_0), \quad \Lambda_3 = a^2(m^2 - E^2).$ (46)

We compare this equation with Eq. (1) and obtain $c_1 = 1$, $c_2 = -2$, $c_3 = -1$, $q_1 = \sqrt{\Lambda_3}$, $p_1 = \sqrt{H}$ where $H$ is defined by Eq. (7). For the energy eigenvalues we use Eq. (10) and write that equation as

$$X^2 + (2n+1)X + n(n+1) = \Lambda_1 \tag{47}$$

where we have defined $X$ as $X = q_1 - p_1 = \sqrt{\Lambda_3} - \sqrt{H}$.

The solution of this quadratic equation gives $\sqrt{\Lambda_3} = \sqrt{H} - (n + 1/2) + \frac{1}{2}\sqrt{1 + 4\Lambda_1}$. Replacing the parameters we get the following equation for the energy eigenvalues

$$E_n^2 - m^2 = -\frac{1}{a^2}\{(n + \frac{1}{2} - a[m^2 - E_n^2 + S_0^2 - V_0^2 - 2(E_n V_0 + mS_0)]^{\frac{1}{2}} - \frac{1}{2}[1 + 4(V_0^2 - S_0^2)]^{\frac{1}{2}}\}^2 \tag{48}$$

The corresponding wave functions are given by Eq. (9) and can be expressed as

$$u_n = N(1-s)^{-\sqrt{H}} s^{\sqrt{\Lambda_3}} P_n^{(2\sqrt{\Lambda_3}, -2\sqrt{H})}(1-2s)$$

To study the K-G equation for any-$\ell$ state the centrifugal term must be added to Eq. (45). In order to find analytical solutions we use the approximate expansion [47-50] $(1/r^2) \cong (1/R^2)(D_0 + D_1 s + D_2 s^2)$ where $D_i$ are some constants and $s$ is defined before. This addition of the centrifugal term changes only the constants $\Lambda_i$ in Eq.(46). If we indicate the new constants by $\Gamma_i$ we find that

$\Gamma_1 = \Lambda_1 + \ell(\ell+1)a^2 D_2 / R^2$, $\Gamma_2 = \Lambda_2 - \ell(\ell+1)a^2 D_1 / R^2$, $\Gamma_3 = \Lambda_3 + \ell(\ell+1)a^2 D_0 / R^2$. The equation for the energy levels becomes

$$E_n^2 - m^2 = -\frac{1}{a^2}\{(n + \frac{1}{2} - a[m^2 - E_n^2 + S_0^2 - V_0^2 - 2(E_n V_0 + mS_0) + \frac{\ell(\ell+1)a^2}{R^2}(D_0 + D_1 + D_2)]^{\frac{1}{2}}$$
$$-\frac{1}{2}[1 + 4(V_0^2 - S_0^2) + 4\frac{\ell(\ell+1)a^2 D_2}{R^2}]^{\frac{1}{2}}\}^2 - \ell(\ell+1)a^2 D_0 / R^2 \tag{49}$$

For the corresponding wave functions only the $\Lambda_3$ and $H$ must be replaced with their new values.

Case 7 : Generalized Pöschl-Teller potential

The Generalized Pöschl-Teller potential [53-59] is defined as

$$V(r) = -V_1 \frac{1}{\cosh^2(\alpha r)} + V_2 \frac{1}{\sinh^2(\alpha r)} \tag{50}$$

We will take this as the vector potential and assume that the scalar potential is equal to vector potential. First we define u(r)=R(r)r and insert this into Eq.(21) which gives a differential equation for u . Then we transform to a new variable $s = \cosh^2(\alpha r)$ and arrive at the following equation

$$\frac{d^2 u}{ds^2} + \frac{(\frac{1}{2} - s)}{s(1-s)} \frac{du}{ds} + \frac{1}{s^2(1-s)^2}[-\Lambda_1 s^2 + \Lambda_2 s + \Lambda_3]u = 0 \tag{51}$$

where $\Lambda_1 = \frac{m^2 - E^2}{4\alpha^2}$, $\Lambda_2 = \frac{m^2 - E^2}{4\alpha^2} - \frac{(E+m)}{2\alpha^2}(V_2 - V_1)$, $\Lambda_3 = \frac{(E+m)V_1}{2\alpha^2}$. Here it is assumed that the scalar potential is equal to vector potential and angular momentum is zero. Comparing this with Eq. (1) we get $c_1 = 1/2$, $c_2 = -1$, $c_3 = -1$ and using Eq.(6,7) we find

$$q_1 = \frac{1}{4}(\sqrt{1 + 16\frac{(E+m)}{2\alpha^2}V_1} + 1), \quad p_1 = \frac{1}{4}(\sqrt{1 + \frac{8}{\alpha^2}(E+m)V_2} - 1), \quad \eta = \frac{1}{2}\sqrt{1 + 8\frac{(E+m)V_1}{\alpha^2}}, \tag{52}$$

$$\beta = -\frac{1}{2}\sqrt{1 + \frac{8}{\alpha^2}(E+m)V_2}.$$

The equation for the energy levels follows from Eq.(10) and the wave functions are given by Eq.(9). The results of this equation can be written as

$$\sqrt{\frac{(m^2 - E_n^2)}{4\alpha^2}} = (n + \frac{1}{2}) - \frac{1}{4}\sqrt{1 + \frac{8}{\alpha^2}(E_n + m)V_2} + \frac{1}{4}\sqrt{1 + \frac{8}{\alpha^2}(E_n + m)V_1} \tag{53}$$

$$u(s) = (1-s)^{-p_1} s^{q_1} P_n^{(\eta,\beta)}(1 - 2s) . \tag{54}$$

To add the centrifugal term we use the approximation [59,60] $1/r^2 \approx \alpha^2/\sinh^2(\alpha r)$. This additional term changes only $\Lambda_2$ to $\Lambda_2 - \ell(\ell+1)/4$ in Eq. (51). This replacement must be done in Eq.(52) and in Eq. (53) for the energy levels and wave functions. We can compare these results with Ref. (57). They have the same potential and they also use an approximation for the centrifugal term. There is an additional constant term in their approximation of the centrifugal term. When that constant is taken as zero, their results are also given by our Eq. (53). That is, our results for this application are consistent with Ref. (57).

## 2  Conclusions

In this work we have given a scheme for the analytic solutions of the Klein-Gordon equation with vector and scalar potentials. We have shown that if the wave equation can be transformed into a differential equation of a certain form it has analytical solutions and these solutions can be written down easily. For several potentials we have obtained the wave function and determined equations for the energy eigenvalues directly. In order to obtain analytic solutions for the non-zero angular momentum case we have used some approximate form of the centrifugal term.